\documentclass[12pt]{article}

\usepackage{mathtools}

\usepackage{multicol}

\usepackage{mathtools}
\usepackage{type1cm}
\usepackage{extsizes}
\usepackage{amsmath,amsxtra,amssymb,latexsym,amscd,amsthm}
\usepackage[a4paper,left=2.0cm,right=2.0cm,top=2.0cm,bottom=3.0cm]{geometry}
\usepackage{parskip}

\usepackage{amssymb}

\begin{document}
\title{\textbf{Mass creation from extra-dimensions}}
\author{ \itshape {Dao Vong Duc\textsuperscript{1}, Nguyen Mong Giao\textsuperscript{2}}
\\ \itshape {$^1$Institute of Physics, Hanoi}
\\ \itshape {$^2$Hung Vuong University, Ho Chi Minh}}
\date{ }
\maketitle



\begin{abstract}
In this work we consider a mechanism for mass creation based on the periodicity condition dicated from the compactification of extradimensions. It is also shown that the existence of Tachyon having negative square mass is closely related to time-like extradimensions. \par
\end{abstract}


\setlength{\parskip}{0pt}

\setlength{\parindent}{0cm} 
\textbf{1. Introduction} \par
\setlength{\parindent}{0.5cm} 
The existence of space-time extradimensions has been a subject of intensive research study during the last decades $[1-3]$.\par
The Topology of extradimension, especially their compactification play a crucial role in many physical aspects, mostly in the construction of various models of Unified theory of interactions, such as Superstring theory, Extended General Ralativity, and so on $[4-7]$. \par 
It is worth noting, on the other hand, that is such approaches the particles mass always remains a problem of actual characters.\par
In this work we propose a mechanism for mass creation through the compactification of space-time extradimensions. The crucial argument is the proposed periodicity condition dictated from the compactification of extradimensions.\par
The original field functions depend on all space-time coordinate components including those for extradimensions, the ordinary field functions ordinary 4-dimensional space-time are considered as effective field functions obtained by integration of the original ones over extra space-time.\par
In section 2, we present some general principles related to the compactification of extradimensions.\par
In section 3, a mechanism for mass creation is treated.\par




\setlength{\parindent}{0cm} 
\textbf{2. Periodicity campactification condition} \par
\setlength{\parindent}{0.5cm} 
For simplicity let us begin with the case of one extra dimension. Denote the 5-dimensional coordinate vector by $x^M$ with $M=\mu,5$. The Greek indices $\mu$, $\nu$,... will be use as conventional 4-dimensional Lorentz indices (0,1,2, and 3). We do not directly care from the extra dimensions is topologically compactified, but instead a specific periodicity condition is put on the field functions depending on extra dimensions, namely\\
\begin{equation} \label{int:MG1}
F(x^\mu,x^\varsigma + L)=f_L(L).F(x^\mu,x^\varsigma)
\end{equation} \par
Where $f_L(L)$ is some parameter function depending on the compactification lenth L. \par
The condition (\ref{int:MG1}) corresponds to the equation:
\begin{equation} \label{int:MG2}
\frac{\partial}{\partial x^\varsigma}F(x^M)=g_L(L).F(x^M)
\end{equation} \par
With the relations:
\begin{equation} \label{int:MG3}
f_L(L) = e^{lgM(L)}
\end{equation}
\begin{center}
$g_L(L) = \frac{1}{L}\left[ln f_L(L)+2 \pi ni \right] (n \in Z ) $
\end{center} \par
In general we can put
\begin{equation} \label{int:MG4}
f_L(L) = \rho_F(L).e^{i \theta_F(L)}
\end{equation}
\begin{center}
$g_L(L) = \frac{1}{L}\left[ln \rho_F(L)+ i(\theta_F(L) + 2 \pi n \right] $
\end{center} \par
For neutral field, $ F^+ = F$ , $ f_L(L) $ is to be real and therefore $ \theta_F = 0$, $ n = 0 $ . \par
The periodicity condition (\ref{int:MG1}) can be generalized for the case of arbitrary number of extra dimensions in the following manner. \par
For convenience we denote the extra dimension coordinates $ x^5,x^6, \cdots ,x^{4+d} $ by $ y^\alpha \equiv x^{4 + \alpha, }, \alpha = 1,2, \cdots, d $ and write \\
\begin{equation} \label{int:MG5}
F(x^M) \equiv F\left( x^\mu , y^\alpha \right) \equiv F(x,y)
\end{equation} \par
The periodicity condition (\ref{int:MG1}) is now generalized to be: \\
\begin{equation} \label{int:MG6}
F\left( x , y^\alpha + L^\alpha \right) = f_F^{(\alpha)}(L^\alpha).F(x,y)
\end{equation}
and the corresponding equation (2) becomes:
\begin{equation} \label{int:MG7}
\frac{\partial}{\partial y^\alpha}F(x,y) = g_F^{(\alpha)}(L_\alpha).F(x,y)
\end{equation}
with the ralations:
\begin{center}
$ f_F^{(\alpha)}(L^\alpha) = f_F^{(\alpha)}(L^\alpha).e^{i \theta_F^{(\alpha)}}(L_\alpha) $
\end{center}
\begin{equation} \label{int:MG8}
 g_F^{(\alpha)}(L_\alpha) = \frac{1}{L^\alpha}\left[ln f_F^{(\alpha)}(L^\alpha) + i( \theta_F^{(\alpha)}(L^\alpha) + 2 \pi n )\right]
\end{equation}
\setlength{\parindent}{0cm}
\textbf{3. Effective field equation and mass} \par
\setlength{\parindent}{0.5cm}
The general procedure of our treatment is as follows. We start from the (4+d) dimensional Lorentz invariant Kinetri Lagrangian L(x,y) and the action for the field F(x,y) defined as
\begin{equation} \label{int:MG9}
S = \int S(y)(dy)
\end{equation}
\begin{center}
$ S(y) \equiv \int d^4 x. L(x,y) $
\end{center} \par
Where $ (dy) \equiv dy^1 . dy^2 ... dy^d$ and the general is performed over the whole extra space time. \par
The principle of minimal action for $ S(y)$ then gives the Euler-Lagrange equation
\begin{equation} \label{int:MG10}
\frac{\partial L(x,y)}{\partial F(x,y)} - \partial_\mu \frac{\partial L(x,y)}{\partial(\partial_\mu . F(x,y))} = 0
\end{equation}  \par
Which in turn leads to the equation of Klein-Gordon type:
\begin{center} 
	$(\Box + m_F^2)F(x) = 0$
\end{center} \par
For the effective field defined as
\begin{equation} \label{int:MG12}
   F(x) \equiv \int (dy) F(x,y)
\end{equation}  \par
For illustration let us consider in more details the cases of scalar, spinor and vector fields. \vskip 2.0cm
\setlength{\parindent}{0.5cm}
\textbf{ \textsl{3.1. Scalar field}} \par
\setlength{\parindent}{1.0cm}
The free neutral scalar field $\Phi(x,y)$ is described by the Lagrangian
\begin{equation} 
\begin{split}
	\L(x,y) &= \frac{1}{2} \partial^M \Phi(x,y) . \partial_M \Phi(x,y) \\
	        & \equiv \frac{1}{2} \left\{ \partial^M \Phi(x,y) \partial_M \Phi(x,y) 
						+ \sum_{ a = 1}^d \varsigma_{aa} \partial_a \Phi(x,y).\partial_a \Phi(x,y) \right\}
 \end{split}
\end{equation} \par
Where $ \partial_a \equiv \frac{\partial}{\partial y^a } , \varsigma_{ab} $ is a MinKonski metric for extra dimensions:
\begin{equation} 
  \notag
	\varsigma_{ab} = 
	    \begin{cases}
			   0, & if a \neq b \\
				 1, & if a = b - timelike \\
				-1, & if a = b - spacelike
			\end{cases}
\end{equation} \par
By inverting (7) into (12), we obtain:
\begin{equation} 
	L(x,y) = \frac{1}{2} \left\{ \partial^M \Phi(x,y) . \partial_M \Phi(x,y)
	+ \sum_{ a = 1}^d \varsigma_{aa} \left( g^{(a)} (L_a)\right)^2 . \Phi^2(x,y) \right\}
\end{equation} \par
And from here the equation
\begin{equation}  
	( \Box + m_\Phi^2 ) \Phi(x) = 0
\end{equation} \par
For the effective field
\begin{equation}  
  \notag
	\Phi (x) = \int (dy) \Phi (x,y)
\end{equation} \par
With
\begin{equation}  
	m_\Phi^2 = - \Sigma_a \varsigma_{aa} \left( g^{(a)} (L_a) \right)^2
\end{equation} \par
It is worth nothing that the squared mass $ m_\phi^2 $ is positive if all the extra dimensions are space-live, and can be negative if there exists time-live extra dimensions. \par
For changed scalar field instead of (12) we take
\begin{equation}
\begin{split}  
L(x,y) &= \partial^M \Phi^+ (x,y). \partial_M \Phi(x,y) \\
       &= \partial^\mu \Phi^+ (x,y). \partial_\mu
\Phi (x,y) + \sum_{a=1}^d \partial^a \Phi^+ (x,y). \partial_a \Phi (x,y)
\end{split}
\end{equation} \par
And instead of (13) we have:
\begin{equation}  
\begin{split}
	L(x,y) &= \partial^M \Phi^+ (x,y). \partial_M \Phi(x,y) \\
	       &+ \sum_{a=1}^d \varsigma_{aa} \left| g_\Phi^{(a)}  (L_a) \right|^2 . \Phi^+ (x,y). \Phi (x,y)
\end{split}
\end{equation} \par
And from here the same equation as (14) with:
\begin{equation}  
	m_\Phi^2 = - \Sigma_a \varsigma_{aa} \left| g_\Phi^{(a)}  (L_a) \right|^2 
\end{equation} \vskip 2.0cm

\setlength{\parindent}{0.5cm}
\textbf{\textsl{3.2. Spinor field}} \par
\setlength{\parindent}{1.0cm}


In (4+d)-dimensional space-time, the spinor field is decribed by a $ 2^{\frac{4+d}{2}} $ component function $ \psi_a (x,y) $ with the free Lagrangian
\begin{equation} 
	L(x,y) = \frac{i}{2} \overline{\psi (x,y)} . \Gamma^M \overline{ \partial_M} \psi(x,y) \\
	       \equiv \frac{i}{2} \left\{ \overline{\psi} \Gamma^\mu \overline{\partial_\mu} \psi + \sum_{a=1}^d \overline{\Psi} \Gamma^{a+4} \overline{\partial_a} \psi \right\}
\end{equation} \par

Where $ \Gamma^M $ denote (4+d) Dirac $ 2^{\frac{4+d}{2}} $ x $2^{\frac{4+d}{2}} $ matrices obeying the ant commutation relations:
\begin{equation} 
\left\{ \Gamma^M, \Gamma^\nu \right\} = 2 \varsigma^{\mu \nu}
\end{equation}
\begin{equation} 
\notag
\left\{ \Gamma^M, \Gamma^{4+a} \right\} = 0
\end{equation}
\begin{equation} 
\notag
\left\{ \Gamma^{4+a}, \Gamma^{4+b} \right\} = 2 \varsigma^{ab}
\end{equation}
\begin{equation} 
\notag
\overline{\psi} \equiv \psi + \Gamma^0
\end{equation} \par
By inverting
\begin{equation} 
\partial_a \psi (x,y) = g_\psi^{(a)} L_a . \psi (x,y)
\end{equation}
\begin{equation} 
\notag
\partial_a \overline{\psi (x,y)} = g_\psi^{(a)} L_a . \overline{\psi (x,y)}
\end{equation} \par
Into (19) we obtain:
\begin{equation}  
	L(x,y) = \frac{i}{2} \overline{\psi (x,y)} . \Gamma^M \overline{ \partial_M} \psi(x,y) - Img_\psi^{(a)} (L_a) \overline{\psi} \Gamma^{4+a} \psi
\end{equation} \par
And from here the equation
\begin{equation} 
\left(i \Gamma^M \partial_\mu - \sum_{a=1}^d img_\psi^{(a)} (L_a) . \Gamma^{4+a} \right) \psi (x,y) = 0\end{equation} \par
By acting from the left both sides of this equation by
\begin{equation} 
\notag
i \Gamma^\upsilon \partial_\upsilon - \sum_{b=1}^d img_\psi^{(b)} (L_b) . \Gamma^{4+b} 
\end{equation} \par
And taking into account the relations (20) we have:
\begin{equation} 
\left\{ \Box - \sum_{a=1}^d \varsigma_{aa} \left(img_\psi^{(a)} (L_a) \right)^2 \right\} \psi (x,y) = 0 
\end{equation} \par
And hence
\begin{equation} 
m_\psi^2 = - \sum_a \varsigma_{aa} \left(img_\psi^{(a)} (L_a) \right)^2  
\end{equation} \par
We note that $ m_\psi^2 > 0 $ if all the extra dimensions are space-like, $ m_\psi^2 = 0 $ if all $ g_\psi^{(a)} $ are real, and $ m_\psi^2 $ can be negative if there exists time-like extra dimension.\par
\setlength{\parindent}{0.5cm}
\textbf{\textsl{3.3. Vector field}} \par
\setlength{\parindent}{1.0cm}
We restrict ourselves to the case d=1 and consider the neutral vector field $ V_M (x,y) $ satisfying the periodicity condition
\begin{equation} 
V_M (x,y+L) = f_V (L) . V_M (x,y) 
\end{equation} \par
And in correspondence
\begin{equation} 
\frac{\partial}{\partial_y} V_M (x,y) = g_V (L) . V_M (x,y) 
\end{equation}
\begin{equation}
\notag 
f_V (L) = e^L  g_V (L) 
\end{equation} \par
The free vector field $ V_M (x,y) $ is described by the Lagrangian
\begin{equation}
\begin{split}
\L(x,y) & {}= \frac{-1}{4} F_{MN} F^{MN} \\ 
& {} = \frac{-1}{4} \left( F_{\mu \upsilon} F^{ \mu \upsilon } + 2 F_{ \mu \varsigma} F^{ \mu \varsigma } \right) \\ 
& {} = \frac{-1}{4} F_{\mu \upsilon} F^{ \mu \upsilon } - \frac{1}{2} \varsigma_{\varsigma \varsigma} \left( \partial_\mu V_\varsigma . \partial^\mu V_\varsigma
 + \partial_\varsigma V_\mu . \partial_\varsigma V^\mu - 2 \partial_\mu V_\varsigma . \partial_\varsigma V^\mu \right)
\end{split}
\end{equation}
Where 
\begin{equation}
\notag 
F_{\mu \upsilon } \equiv \partial_\mu V_\upsilon - \partial_\upsilon V_\mu
\end{equation}
\begin{equation}
\notag 
F_{\mu \varsigma } \equiv \partial_\mu V_\varsigma - \partial_\varsigma V_\mu
\end{equation} \par
By inverting (27) into (28) we have:

\begin{equation}
L(x,y) = \frac{-1}{4} F_{\mu \upsilon } F^{\mu \upsilon}  
- \frac{1}{2} \varsigma_{\varsigma\varsigma} \left( \partial_\mu V_\varsigma . \partial^\mu V_\varsigma + g_V^2 (L) V_\mu V^\mu - 2 g_V (L) \partial_\mu V_\varsigma V^\mu \right)
\end{equation} \par
Now we define a new physical vector field $ W_\mu $ by putting
\begin{equation}
W_\mu \equiv V_\mu - \frac{1}{g_V (L)} \partial_\mu V_\varsigma
\end{equation} \par
Expressed in terms of $ W_\mu $, the Lagrangian (29) has the form:
\begin{equation}
L(x,y) = - \frac{1}{4} G_{\mu \upsilon} G^{\mu \upsilon} - \frac{1}{2} \varsigma_{\varsigma \varsigma} g_V^2 (L) W_\mu W^\mu
\end{equation}
\begin{equation}
\notag
G_{\mu \upsilon} \equiv \partial_\mu W_\upsilon - \partial_\upsilon W^\mu
\end{equation} \par
The Lagrangian (31) leads to the equation:
\begin{equation}
\left( \Box - \varsigma_{\varsigma \varsigma } g_V^2 (L) \right)  W_\mu =0
\end{equation} \par
Which means that the effective vector field
\begin{equation}
\notag
W_\mu (x) = \int_0^L dy. W_\mu (x,y)
\end{equation} \par
Has squared means
\begin{equation}
m_W^2 = - \varsigma_{\varsigma \varsigma} g_V^2 (L)
\end{equation} \par
It's positive or negative depending upon whether the extra dimension is space-like or time-like. \\
\setlength{\parindent}{0cm}
\textbf{4. Conclusion} \par
\setlength{\parindent}{0.5cm}
In this work we have proposed a mechanism for the creation of particle mass. The key idea is that the mass is originated from the compactification of extra dimensions followed by the periodicity condition for the particle fields. \par
It is worth nothing that according to the mechanism the existence of tachyon having negative squared mass is closely related to the existence of time-like extra dimensions.
\bibliographystyle{elsarticle-num}



\end{document}